# Fabrication of Surface-Patterned ZnO Thin Films Using Sol-gel Methods and Nanoimprint Lithography

Shuxi Dai, Yang Wang, Dianbo Zhang, Xiao Han, Qing Shi, Shujie Wang, Zuliang Du*

**Abstract**

Surface-patterned ZnO thin films were fabricated by direct imprinting on ZnO sol and subsequent annealing process. The polymer-based ZnO sols were deposited on various substrates for the nanoimprint lithography and converted to surface-patterned ZnO gel films during the thermal curing nanoimprint process. Finally, crystalline ZnO films were obtained by subsequent annealing of the patterned ZnO gel films. The optical characterization indicates that the surface patterning of ZnO thin films can lead to an enhanced transmittance. Large-scale ZnO thin films with different patterns can be fabricated by various easy-made ordered templates using this combination of sol-gel and nanoimprint lithography techniques.

S. Dai, Y. Wang, D. Zhang, X. Han, Q. Shi, S. Wang, Z. Du (corresponding author)

Key Laboratory for Special Functional Materials of Ministry of Education

Henan University, Kaifeng 475004, P.R. China

*Corresponding author. e-mail: zld@henu.edu.cn.

# 1 Introduction

Zinc oxide (ZnO) is a promising semiconductor with a wide band gap of 3.37 eV at room temperature with high mechanical and thermal stabilities. The ZnO thin films have attracted significant attention due to their wide range of electrical and optical properties in recent years [1, 2]. These advantages are of considerable interest for practical applications in electronics and optoelectronics such as solar cells, gas sensors, surface acoustic wave devices and other optoelectronic devices like LED and FPD [3-5]. There are many methods for fabricating ZnO thin films, such as magnetron sputtering [4, 5], thermal evaporation [6], and sol–gel process [7].

For the technical applications, the preparation of functional ZnO structures and electronic devices often demands large-scale surface-patterned ZnO thin films. The microelectronic and optoelectronic technology requires the construction of micro/nano devices with patterned functional materials. It is essential that to design and control the pattern structure, periodicity, interfeature distance and positions to realize new properties for the further nanodevice applications in various areas. Lipowsky et al. reported the preparation of patterned ZnO films by selective deposition on self-assembled monolayer template [8]. Guzman et al. had used ZnO nanoparticles as ink in confined dewetting lithography to obtain patterned ZnO films [9]. Recently, the polymer-based ZnO precursor solutions had been used as resistance layer in soft lithography process of ceramic patterns [10].

However, it is still a challenge to fabricate patterned ZnO thin films over large scale at a cost-effective way for the industrial applications. Nanoimprint lithography (NIL) has been demonstrated as a high-throughput, low cost and nonconventional lithographic technique for the potential applications in electronics and photonics fields. The sol-gel technique offers the possibility of preparing a small as well as large-area coating of ZnO thin films at low cost. The combination of both techniques would give a new way for the large-scale fabrication of surface-patterned ZnO thin films.

In the present study, large-scale ZnO thin films with various patterns were prepared using nanoimprint lithography technique with polymeric ZnO sols deposited on various substrates. The polymeric ZnO sol was imprinted and converted to ZnO gel under the thermal curing nanoimprint process. PDMS soft stamps were used to pattern the ZnO sol-gel films in order to avoid the limitation of hard stamps which is typically time consuming to produce and thus very expensive. Various low-cost and easy-made ordered structures had been replicated with the flexible PDMS stamps. Imprinted ZnO films were prepared with the unique patterns of master molds. The influence of surface patterning on the optical properties was also observed and discussed.

## 2 Experimental

### 2.1 materials

All chemicals were used as received. Poly(dimethylsiloxane) (PDMS) was commercially obtained (Sylgard 184 kits, Dow Corning). Poly(acrylicacid) (PAA, Mw = 1800) was from Aldrich. $Zn(NO_3)_2$ (98.5%), analytical grade 2-methoxyethanol and ethanol were used as received in the experiments. Deionized water (Millipore, USA) with a resistance of 18 MΩ cm$^{-1}$ was used. Various substrates such as pieces of silicon wafer (100), and quartz microscope coverslips were used in experiments. All substrates were cleaned by RCA standard cleaning method [11].

### 2.2 Fabrication of PDMS mold

PDMS were used to replicate ordered structures and then act as stamp in nanoimprint process. Silicon calibration grating (Seiko Co. Ltd.) for AFM measurements, DVD and polystyrene sphere monolayer were used as master molds for the PDMS replication. The PDMS stamp was prepared by casting a 10:1 weight ratio mixture of basic agent and firming agent on the master mold [12, 13]. Then the PDMS stamp was cured at 80 ℃ for 1 h under vacuum to extract air bubbles and replicate feature of master mold. Finally, PDMS stamp was obtained after a complete release from the master mold.

## 2.3 Preparation of ZnO precursor solution

ZnO precursor solution was prepared by dissolving 2.95 mM $Zn(NO_3)_2$ $6H_2O$ in 0.87 ml of 2-methoxyethanol and 0.9 ml of water. Then poly(acrylicacid) (PAA) was added into the solution as a sol stabilizer and dispersant while maintaining the mass ratio of nitrate/PAA at 1:1, and gently stirred until completely melt to obtain ZnO sol. Finally, the ZnO sol can be evenly spin-coated on the various substrates (Si wafer, quartz) for the measurements of various properties of ZnO samples.

## 2.4 The process of nanoimprinting and annealing

Figure 1a shows the main fabrication procedure of the patterned ZnO thin films and Figure 1b shows parameter variation of the nanoimprint process. The imprinting process was done using a commercial nanoimprinter (NIL-2.5, Obducat AB, Sweden) with an optimum two-stage nanoimprint process. In the first stage, polymeric ZnO precursor sol spin-coated on the substrate was pressed with PDMS stamp at a low pressure and low temperature in order to fill the nano-sized grooves of PDMS stamp with sol. In the second stage, the temperature was raised to 150 ℃ with increasing pressure to $2 \times 10^6$ Pa. The ZnO sols on the substrates were converted to stable patterned ZnO gel thin films after several minute thermal nanoimprint process.

Then PDMS stamp was released and the patterned ZnO gel films were heated to yield crystalline ZnO films. The calcination process was from 20 to 550 ℃ with 5 ℃/min heating rate in atmospheric ambient, held there, and cooled with the same rate, to degrade the polymer and remove the degradation products. For control experiments, polymeric ZnO precursor sols were spin-coated onto various substrates at 2500 rpm to form ZnO sol films. Then the ZnO sol films were converted to ZnO gel films by heating of 150 ℃ in air for 20 min. Finally, the ZnO gel full films were annealed to yield ZnO thin films at the same calcination condition mentioned above for imprinted ZnO gel films. The structure and the optical properties of spin coated (non patterned) and imprinted (patterned) ZnO films have been compared.

**2.5 Characterization**

Surface morphology measurements of all films were performed with Seiko SPA400 atomic force microscope (AFM). The crystal structure of films was identified by X-ray diffraction (XRD) using a Philips X-ray diffractometer (X'Pert Pro, Holland) with Cu Kα radiation. The UV-Vis transmittance spectra of ZnO films on quartz microscope coverslips were measured from 300 to 800 nm by a UV-Vis absorption spectroscopy (Lambda35, Perkinelmer US).

**3. Results and discussion**

**3.1 ZnO gel films with line patterns by thermal nanoimprinting.**

Silicon grating templates with line patterns were used as master mold for PDMS to replicate its array structures. Figure 2a and 2b showed the AFM images of the silicon grafting and its replicated PDMS stamp. The silicon grating with a 500 ± 20 nm linewidth and a 35 ± 5 nm height can be seen in Fig. 2a. Measured from the AFM height image in Fig. 2b, the replicated PDMS stamp with line patterns had an average width of 500 ± 50 nm and height of 30 ± 5 nm. It indicated that the PDMS replicated the Si grating template well without obvious defects as compared in the AFM observations.

Patterned ZnO gel films were fabricated by nanoimprint of polymeric ZnO sols with PDMS stamp. There is a typical sol-gel transition during the thermal nanoimprint process, the polymeric precursor ZnO sols were curing by thermal force and organic solvents in the precursor sols were diffused into PDMS polymeric stamps during this thermal curing stage. Figure 2c shows the AFM images of surface-patterned ZnO gel films after nanoimprint process. Measured from the AFM height image, the line array structures had a width of 500 ± 50 nm and a height of 30 ± 2 nm. The linewidth and height of ZnO sol-gel films corresponded well to the linewidth and the height of channels of the PDMS stamp shown in Fig. 2b. Film thicknesses of imprinted ZnO gel film and spin coated ZnO gel film on silicon substrate were estimated by AFM after scratching the films. The average film thickness of imprinted ZnO gel film was 110 nm. It showed that the imprinted ZnO gel films composed of a surface layer of line array structures with the thickness of 30 nm and a bottom layer of full ZnO gel films with thickness of 80 nm. The thickness of spin coated ZnO gel film was also determined as about 110 nm.

### 3.2 Crystalline ZnO films by subsequent annealing

The ZnO gel films were then converted into crystalline ZnO patterns by calcination at 550 ℃ for 1 h. Figure 2d shows the AFM image of ZnO films after annealing. As measured from the AFM height images, the dimensions of ZnO array pattern were reduced to a linewidth of 500 ± 50 nm and a height of 5 ± 1 nm. Compared to the height of surface patterns on ZnO gel films, the dimensions of ZnO patterns were decreased by 80% in the z-direction after calcination. Film thickness measurements of surface-patterned and nonpatterned spin coated ZnO films after calcination were performed by AFM after scratching the films. The average film thickness of surface-patterned ZnO film was 20 nm. It indicated that the surface-patterned ZnO film composed of a 5 nm surface-patterned layer and a 15 nm full ZnO bottom layer. The thickness of nonpatterned spin coated ZnO film was reduced to 20 nm from 110 nm after calcination.

Figure 3 shows the XRD patterns of (line a) surface-patterned and (line b)

nonpatterned spin coated ZnO films after calcinations. The XRD patterns correspond to three main diffraction peaks of crystallized ZnO, namely (100), (002) and (101). It revealed that both kinds of calcined ZnO films were polycrystalline phase with a hexagonal wurtzite structure (Zincite, JCPDS 36-1451) [14-17]. For the surface-patterned ZnO films, the intensity of (002) peak was weaker compared with that of spin coated ZnO films. These diffraction patterns indicated that the intensities of diffraction peaks declined as the patterned structures were formed, which caused the crystallinity to degenerate. Both kinds of patterned and nonpatterned ZnO films were prepared from the same ZnO sol by the spin-coating method and calcined in the same annealing condition. The only difference of two films is the formation process of ZnO gel films with separate preheating condition.

For patterned ZnO films, the gel film was formed during the imprint process at 150 ℃ in an enclosed space inside the nanoimprint lithography equipment. The non-patterned spin coated ZnO gel film was obtained in a muffle furnace by heating of 150 ℃ in the ambient atmosphere. The solvent vaporization and polymer decomposition of non-patterned ZnO gel film can take place much better for the preheating in air. The spin coated ZnO gel film was given an enough condition to remove the organic solvents before crystallization. For the imprinted ZnO film, it was hard to remove all the solvents and polymer inside the gel film before the calcination. The structural relaxation of the imprinted ZnO gel film induced by the solvent vaporization and polymer decomposition, can take place during the subsequent crystallization process. It will greatly influence the crystallinity of final calcined imprinted ZnO films and yield less crystalline films as the result shown in figure 3 and correspond to some reported references [18-19].

**3.3 Enhanced transmittance of surface-patterned ZnO films.**

Figure 4 shows the UV-Vis transmittance spectra of surface-patterned (line a) and nonpatterned spin coated ZnO films (line b) on quartz substrates. In the visible light region, the transmittance is not significantly enhanced by the surface-patterned ZnO films compared with the nonpatterned spin coated ZnO films. The average transmittance value of patterned (line a) ZnO films and nonpatterned (line b) is 85 and 75% in the visible range (400 - 800 nm), respectively. The transmittance values of two kinds of ZnO films decrease quickly with the decrease of wavelength into the UV region (300 - 400 nm). However, the average transmittance value of patterned ZnO films is twice enhanced than that of nonpatterned full ZnO films in the range of 300 - 360 nm. For example, the transmittance value of nonpatterned full films and patterned ZnO films is 25 and 55% at the wavelength of 300 nm, respectively. From above AFM thickness measurement results, we found that the imprinted and spin coated ZnO full films had almost the same thickness of 20 nm. However, the patterned ZnO film composed of a 5 nm top layer of line array structures of about 50% surface area coverage and a 15 nm bottom layer of full ZnO film. So the only structural difference for the two kinds of ZnO is the 5 nm top layer of line array structures in the surface-patterned ZnO films. The top ordered line array structures of patterned ZnO with 50% surface area coverage can effectively reduce the optical path and suppress the reflectance of incident light. The transmittance results in Fig. 4 clearly show that the line arrays structures results in an improved transmission for the patterned ZnO films compared to the nonpatterned full films annealed under the same conditions.

**3.4 Large-scale ZnO thin films with different patterns.**

The nanoimprint stamps are usually fabricated via electron-beam lithography or focused ion beam approach [20-21]. The time-consuming and expensive stamp fabrication process of nanoimprint lithography limits its large-scale applications. Some micro- and nanostructures from nature such as lotus leaf and insect wings were used as cheap stamps [22-23]. In present work, PDMS was used to replicate various low-cost and easy-made ordered structures and then acted as stamp in nanoimprint process. Then ZnO films with different patterns were prepared by the nanoimprint lithography and sol-gel methods.

Figure 5 shows the surface morphologies of the each step in nanoimprint process including the master mold, replicated PDMS stamp and the imprinted ZnO patterns. These typical nanoimprint processes using low-cost DVD and polystyrene sphere monolayer as master mold. In the AFM image of Fig. 5a, the surface of DVD mold showed irregular island-like structures with submicron linewidth. The islands had the length ranged from 900 nm to 1.7 um and an average height of 145 nm. Figure 5b presents the surface of PDMS stamp with negative replicated structures agree well with the original surface structures of DVD. Figure 5c shows the surface structures of imprinted ZnO films with a height of 140 nm. It can be seen that it is a perfect duplicate of the original DVD patterns.

Polystyrene sphere (PS) monolayer was usually used as template for subsequent deposition of metals and semiconductor nanoparticles [24-25]. Here, PDMS were used to replicate the long range ordered structures of PS monolayer and act as nanoimprint stamp. Figure 5d shows the AFM images of the PS monolayer self-assembled on silicon substrates. It has a large-scale highly ordered array structures with a hexagonal arrangement. Figure 5e presents the surface of PDMS soft stamp with negative replicated structure of PS monolayer. The ordered dot-like arrays structures of ZnO thin films can be obtained by the nanoimprint of PDMS stamp on ZnO sol. The hexagonal arrangement of the aligned ZnO dots with a diameter of about 250 nm can be clearly distinguished in Fig. 5f. Other unique natural surface structures like the lotus leaves and insect wings can also be replicated by PDMS stamp and imprinted onto the ZnO sols.

The chemical and physical properties of these surface-patterned ZnO thin films are under investigation.

## 4. Conclusion

Large-scale fabrication of patterned ZnO thin films can be achieved by the combination of nanoimprint lithography and sol-gel techniques. Patterned ZnO thin films were fabricated by direct imprinting of polymer-based ZnO sols deposited on various substrates. The imprinted ZnO gel films were subsequently annealed to form patterned crystalline ZnO thin films. The lower surface area coverage of patterned ZnO films leads to enhanced transmission properties. Using PDMS to replicate different templates and act as nanoimprint stamps, we had successfully fabricated large-area patterned ZnO thin films with various surface structures. The described surface patterning method is simple and suitable to be extended to large-scale processing and other sol-gel derived materials.

**Acknowledgments**

This work was supported by the National Natural Science Foundation of China (Grant No. 20903034, 10874040), the Cultivation Fund of the Key Scientific and Technical Innovation Project, Ministry of Education of China (Grant No. 708062) and Young Backbone Teacher Cultivating Project of Henan University.

**Figure captions**

Fig. 1 (a) Schematic procedure of ZnO patterned films using nanoimprint and annealing and (b) process flow diagram of nanoimprint with temperature vs time (line a) and pressure vs time (line b) plots.

Fig. 2 AFM images of (a) silicon grating (b) PDMS soft stamp (c) imprinted ZnO gel films (d) patterned ZnO films after calcination.

Fig. 3 X-ray diffraction patterns of crystalline (a) patterned ZnO films and (b) nonpatterned ZnO films on silicon substrates after calcinations.

Fig. 4 UV-Vis transmittance spectra of (a) patterned ZnO film and (b) nonpatterned ZnO film on quartz coverslips.

Fig. 5 AFM images of (a) blue-ray DVD and (b) PDMS stamp with negative replicated structures of DVD, (c) imprinted ZnO patterned film with DVD structures, (d) PS monolayer , (e) PDMS stamp with negative replicated structures of PS monolayer, (f) imprinted ZnO patterned film with dot-like arrangement.

**Fig.1**

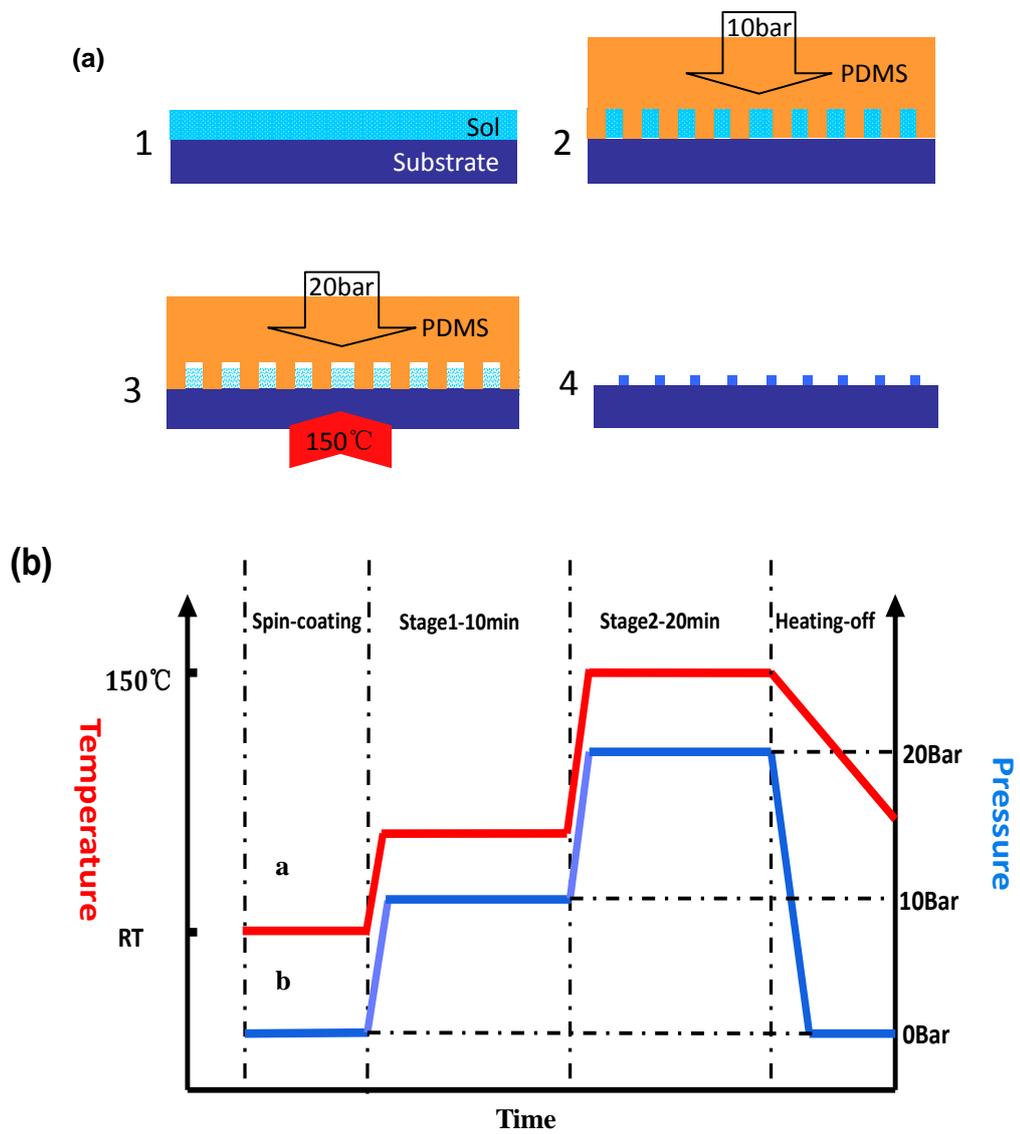

Fig. 1 (a) Schematic procedure of ZnO patterned films using nanoimprint and annealing and (b) process flow diagram of nanoimprint with temperature vs time (line a) and pressure vs time (line b) plots.

**Fig.2**

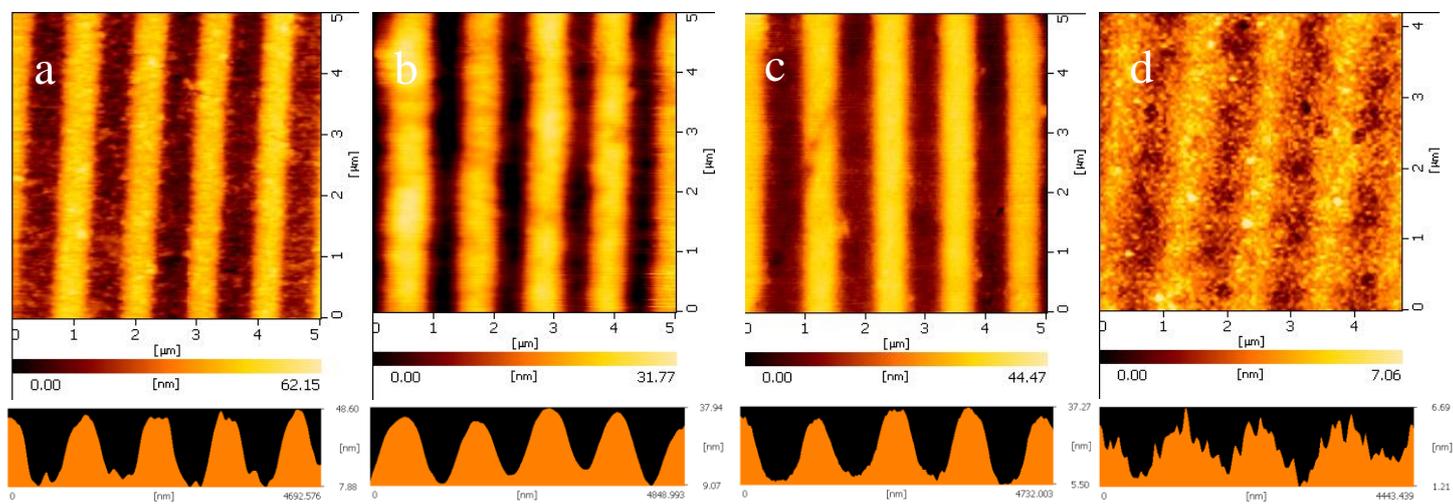

Fig. 2 AFM images of (a) silicon grating (b) PDMS soft stamp (c) imprinted ZnO gel films (d) patterned ZnO films after calcination.

**Fig.3**

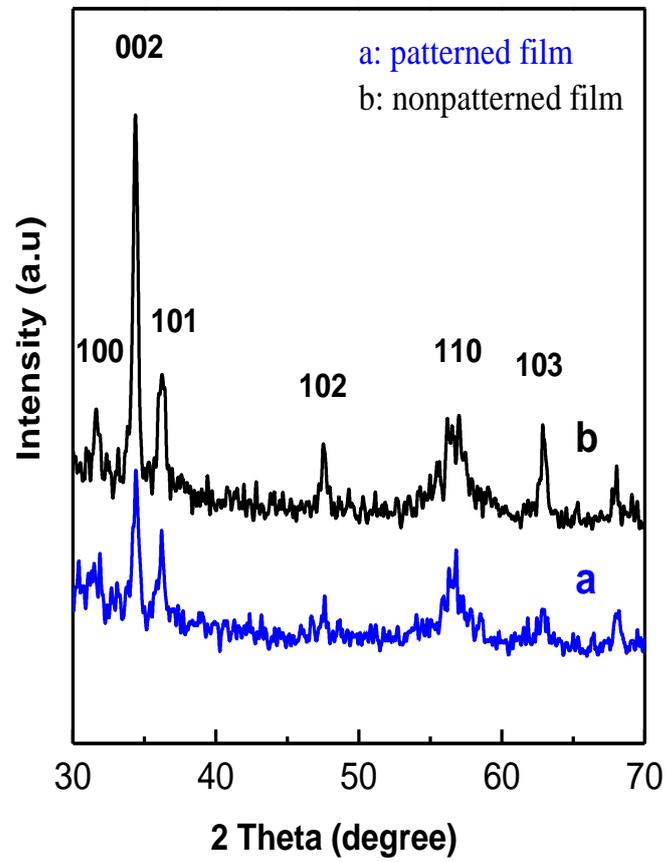

Fig. 3 X-ray diffraction patterns of crystalline (a) patterned ZnO films and (b) nonpatterned ZnO films on silicon substrates after calcinations.

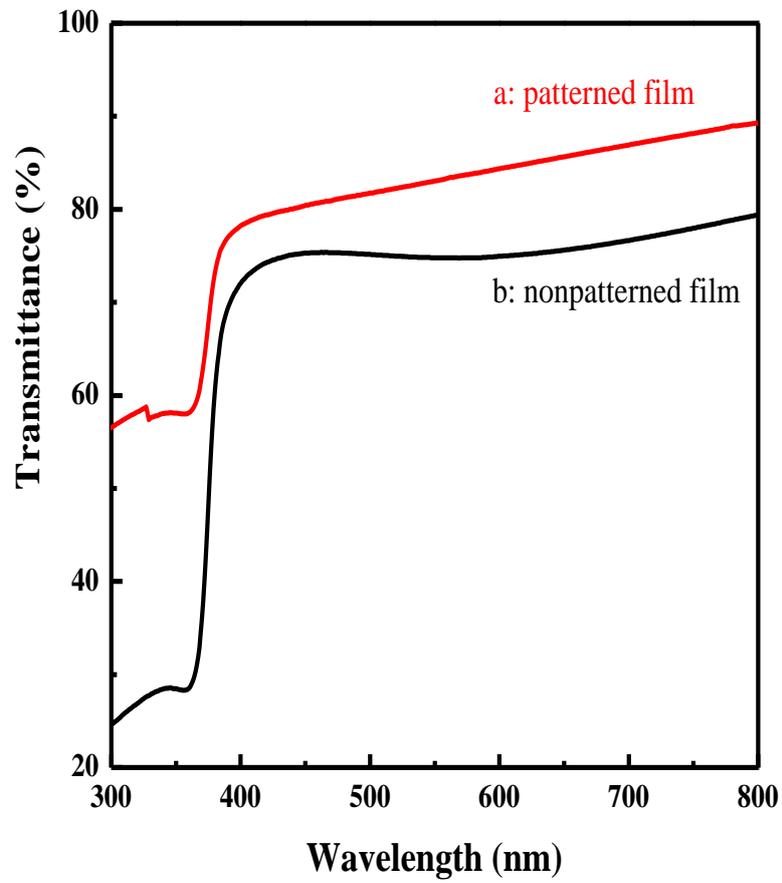

**Fig.4**

Fig. 4 UV-Vis transmittance spectra of (a) patterned ZnO film and (b) nonpatterned ZnO film on quartz coverslips.

**Fig.5**

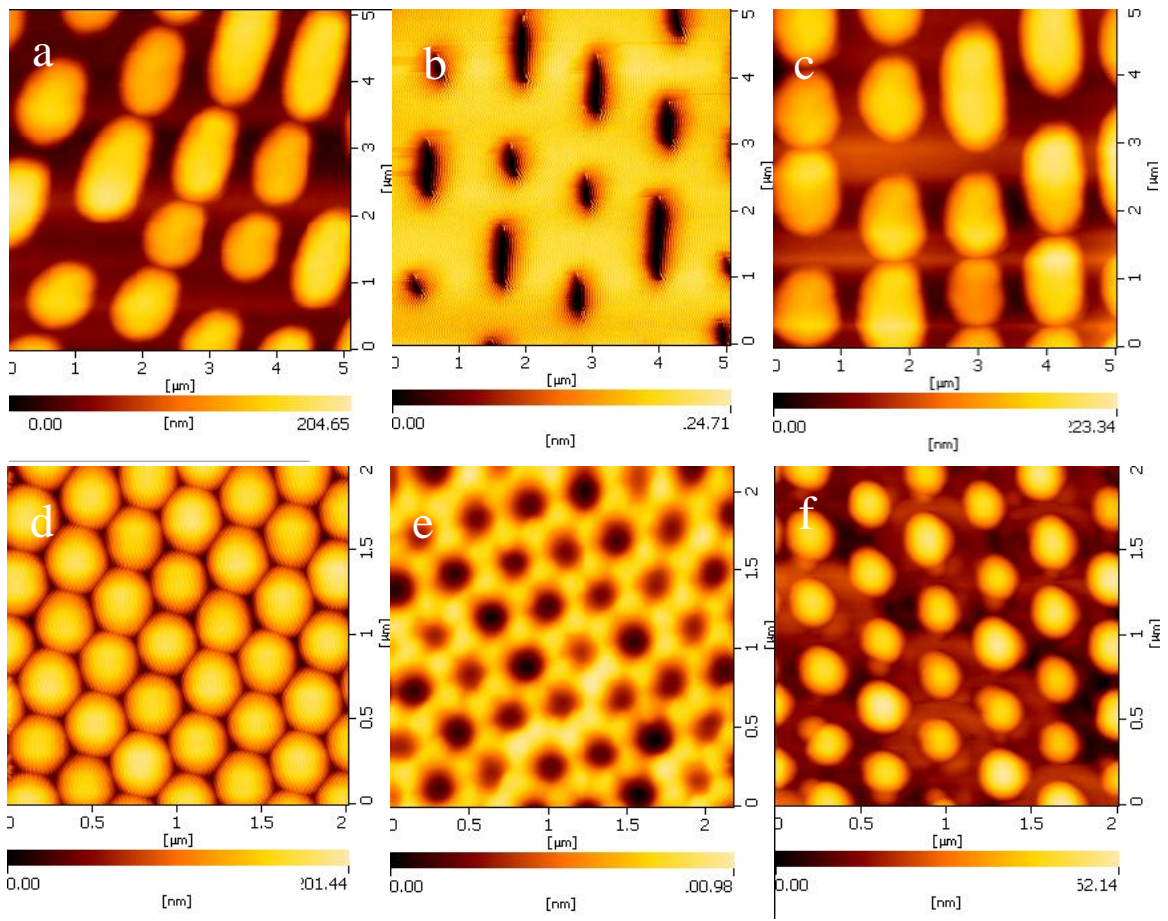

Fig. 5 AFM images of (a) blue-ray DVD and (b) PDMS stamp with negative replicated structures of DVD, (c) imprinted ZnO patterned film with DVD structures, (d) PS monolayer , (e) PDMS stamp with negative replicated structures of PS monolayer, (f) imprinted ZnO patterned film with dot-like arrangement.